\input{aipcheck}

\documentclass[
    ,final            
  ]
  {aipproc}

\layoutstyle{6x9}

\begin{document}

\title{Constraints on $\Delta$G through Longitudinal Double Spin Asymmetry
 Measurements of Inclusive Jet Production in Polarized p+p Collisions at 200
 GeV}


\author{M. Sarsour\footnote{Present address:  Department of Physics and
 Astronomy, Georgia State University, Atlanta, GA 30303.}~ for the STAR
 Collaboration}{
  address={Cyclotron Institute, Texas A$\&$M University, College Station,
 TX 77843}
}

\begin{abstract}
 We report measurements of the longitudinal double spin asymmetry for
 inclusive jet production using polarized p+p collisions at $\sqrt{s} = 200$
 GeV at RHIC. Results from the 2005 and 2006 runs are presented. These results
 set substantial new constraints on the polarized gluon distribution in the
 proton over the kinematic range $0.02 < x< 0.3$, when compared to
 next-to-leading order global analyses of DIS data. The first measurement of
 the transverse single-spin asymmetry for inclusive jets at mid-rapidity is
 also presented.
\end{abstract}

\maketitle


\section{INTRODUCTION}
  Polarized p+p collisions at the Relativistic Heavy Ion Collider (RHIC)
 provide a very suitable environment, rich with strongly interacting probes,
 to constrain $\Delta G$, the integral of the gluon polarized distribution function,
 $\Delta g(x,Q^2)$, evaluated
 at the input scale $Q_0^2$ = 0.4 GeV$^2$, directly and more
 precisely than previously attained. There are many processes where the gluon
 participates directly, and the high center of mass energy, $\sqrt{s} = 200$
 GeV, and high transverse momentum, $p_T$, make NLO pQCD analysis more
 reliable. Indeed, several unpolarized p+p cross sections for reactions
 sensitive to gluons have already been measured at RHIC and are described
 well by pQCD predictions~\cite{PRL97_252001}. The spin program of the
 Solenoid Tracker at RHIC (STAR) experiment~\cite{NIMA499_624} aims, in the
 short term, to utilize these advantages to constrain $\Delta G$ over the
 kinematic range $0.02 < x < 0.3$.  Coincidence measurements can provide the
 $x$ dependence of $\Delta G$, but as RHIC was still developing higher
 luminosity and polarization, the abundant channels for inclusive jet and
 pion production were exploited. These proceedings report
 on the inclusive
 jet production asymmetry, which is an excellent channel to study the gluon
 polarization due to its large cross section and relative independence from
 fragmentation functions. The longitudinal double-spin asymmetry, $A_{LL}$,
 for inclusive jet production is defined as,
\begin{equation}
A_{LL}=\frac{\sigma^{++}-\sigma^{+-}}{\sigma^{++}+\sigma^{+-}}
\end{equation}
where $\sigma^{++}$ ($\sigma^{+-}$) is the inclusive jet cross section when
 the colliding proton beams have equal (opposite) helicities.
 The inclusive measurements from RHIC provide significant
 constraints on $\Delta G$ in the accessible kinematic range.

\section{EXPERIMENT AND ANALYSIS}

   The STAR detector subsystems~\cite{NIMA499_624} used in this measurement
 include the Time Projection Chamber (TPC), the Barrel (BEMC) and EndCap
 (EEMC) electromagnetic calorimeters and the Beam-Beam Counters (BBC). The
 TPC with pseudorapidity $|\eta| < 1.3$ and full azimuthal coverage
 is used to determine the momentum of charged particles. The BEMC and EEMC
 are lead-scintillator sampling calorimeters and have full azimuthal coverage
 spanning the pseudorapidities $|\eta| < 1.0$ and
 $1.08 < \eta < 2.0$, respectively. The calorimeters provide triggering and
 detection of photons and electrons. In 2005 only the west side of the BEMC,
 $0.0 < \eta < 1.0$, was active and the EEMC information was not included
 in the analysis. The BBC is mounted around the beam at longitudinal
 positions $z = \pm 370$ cm, with  full azimuthal coverage and pseudorapidity
 $3.3 < |\eta| < 5.0$. It was used for triggering, luminosity measurement, and
 local polarimetry to determine non-longitudinal polarization components.

  Jets were reconstructed using the midpoint cone
 algorithm~\cite{hep-ex/0005012} which clusters TPC charged track momenta and
 BEMC (and EEMC in 2006) tower energy deposits within a cone radius
 $R=\sqrt{\Delta \eta^2 + \Delta \phi^2}$. Only reconstructed TPC tracks and
 BEMC energy deposits above 0.2 GeV were used by the algorithm. The energy
 seed threshold was 0.5 GeV and jets were merged if more than $50\%$ of their
 energy was of common origin. A minimum jet $p_T=5$ GeV/c was also required.
 The jet axis was required to be within a fiducial range $0.2 < \eta < 0.8$
 ($-0.7 < \eta < 0.9$) for 2005 (2006) data combined with a cone radius of
 $R$ = 0.4(0.7) to reduce the edge effects of the calorimeter acceptance.
 In addition, the BBC time information was used to select events with a
 reconstructed vertex on the beam axis within approximately $\pm$ 60 cm from
 the TPC center to ensure a uniform tracking efficiency.
 Data were collected both with a high tower trigger (HT) that required a minimum
 energy deposition in a $\Delta \eta \times \Delta \phi = 0.05 \times 0.05$
 tower and with a jet patch trigger (JP) that required a minimum energy deposition
 over a $\Delta \eta \times \Delta \phi = 1 \times 1$ region, in coincidence with
 the minimum bias trigger condition. JP triggers were dominant in both 2005 and 2006,
 and in this presentation we show only JP triggers from 2006 data.

\section{RESULTS AND SUMMARY}

   In 2005 we had an order of magnitude increase in the figure of merit over
 our previous measurements \cite{PRL97_252001} with coverage in jet $p_T$ up to 30 GeV/c. This
 allowed us to perform a quantitative comparison of our measured results~\cite{PRL100_232003} to
 global fits of polarized deep-inelastic scattering (DIS) data within the
 GRSV framework \cite{PRD63_094005} for various fixed values of $\Delta G$ \cite{MSnWV}.
 These comparisons showed that the global fits that predict very large
 $\Delta G$ values and large
 $A_{LL}$ are excluded~\cite{PRL100_232003}. However, some fits like GS-C~\cite{PRD53_6100}
 which has a large positive gluon polarization at low $x$, a node near
 $x \sim 0.1$, and a negative gluon polarization at large $x$ are still
 consistent with the data.

\begin{figure}[htp!]
  \includegraphics[scale=0.6]{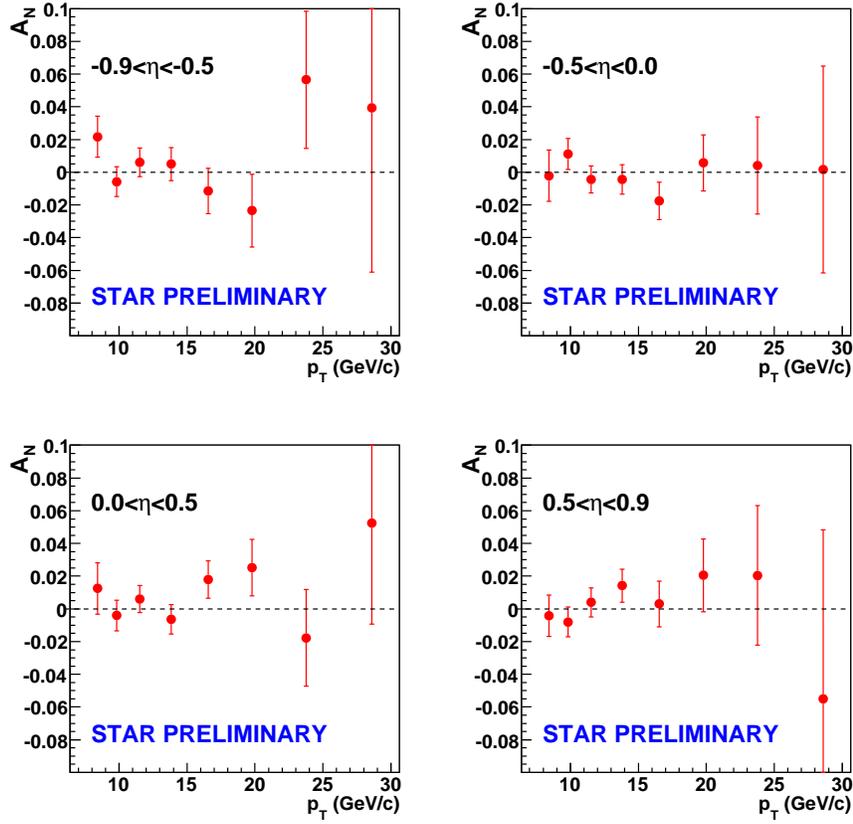}
  \caption{\label{fig:aNvspT}\small Preliminary 2006 $A_N$ for inclusive
 jet production at $\sqrt{s} = 200$ GeV versus jet $p_T$ for four different
 $\eta$ bins.}
\end{figure}

 In 2006, we performed the first measurements of the transverse single-spin
 asymmetry, $A_N$, for inclusive jet production at mid-rapidity.
 Figure~\ref{fig:aNvspT} shows $A_N$ for four different pseudorapidity ranges
 covering the calorimeter acceptance. $A_N$ is consistent with zero within
 the statistical uncertainties.

\begin{figure}[htp!]
 \includegraphics[scale=0.39]{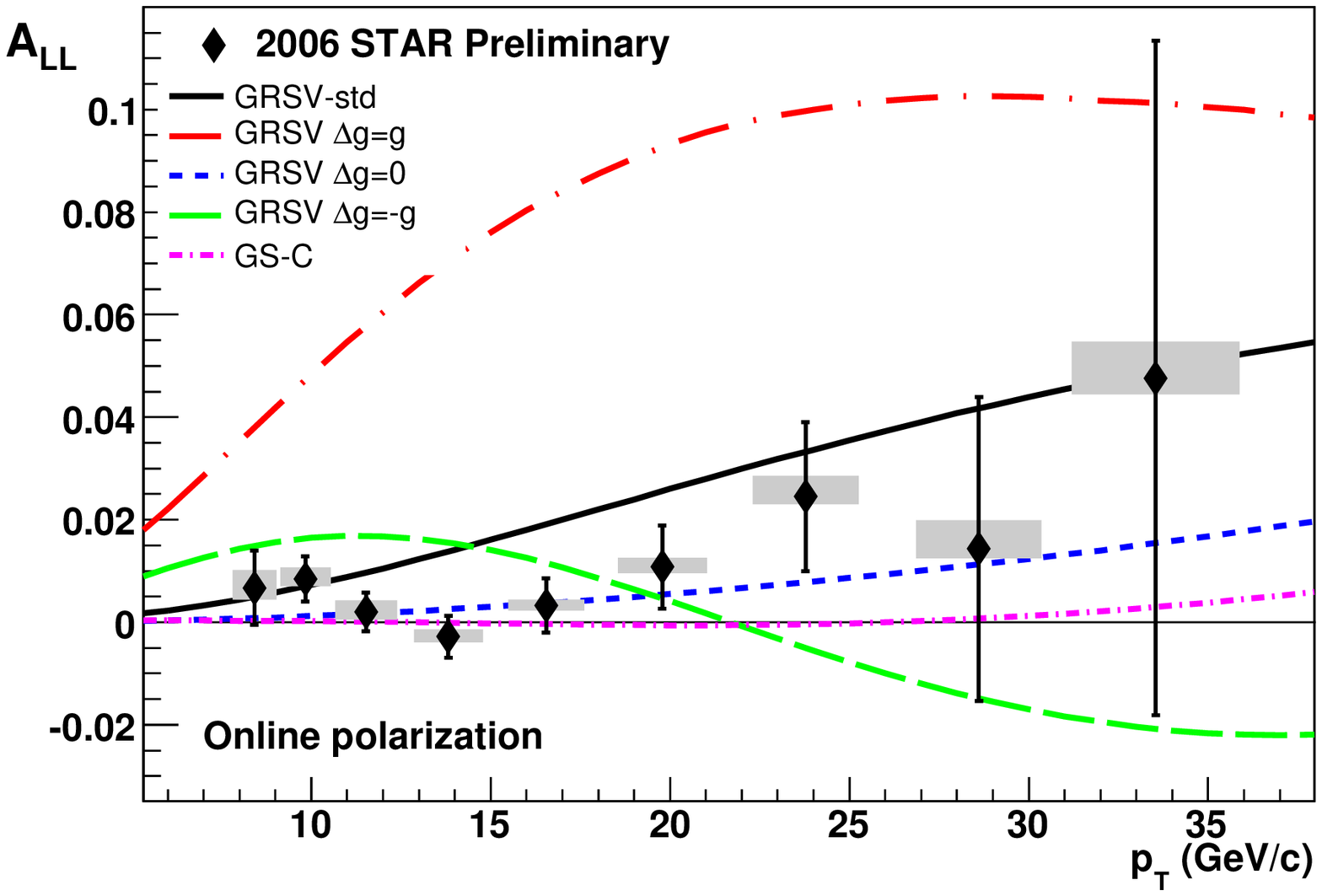}
 \includegraphics[scale=0.348]{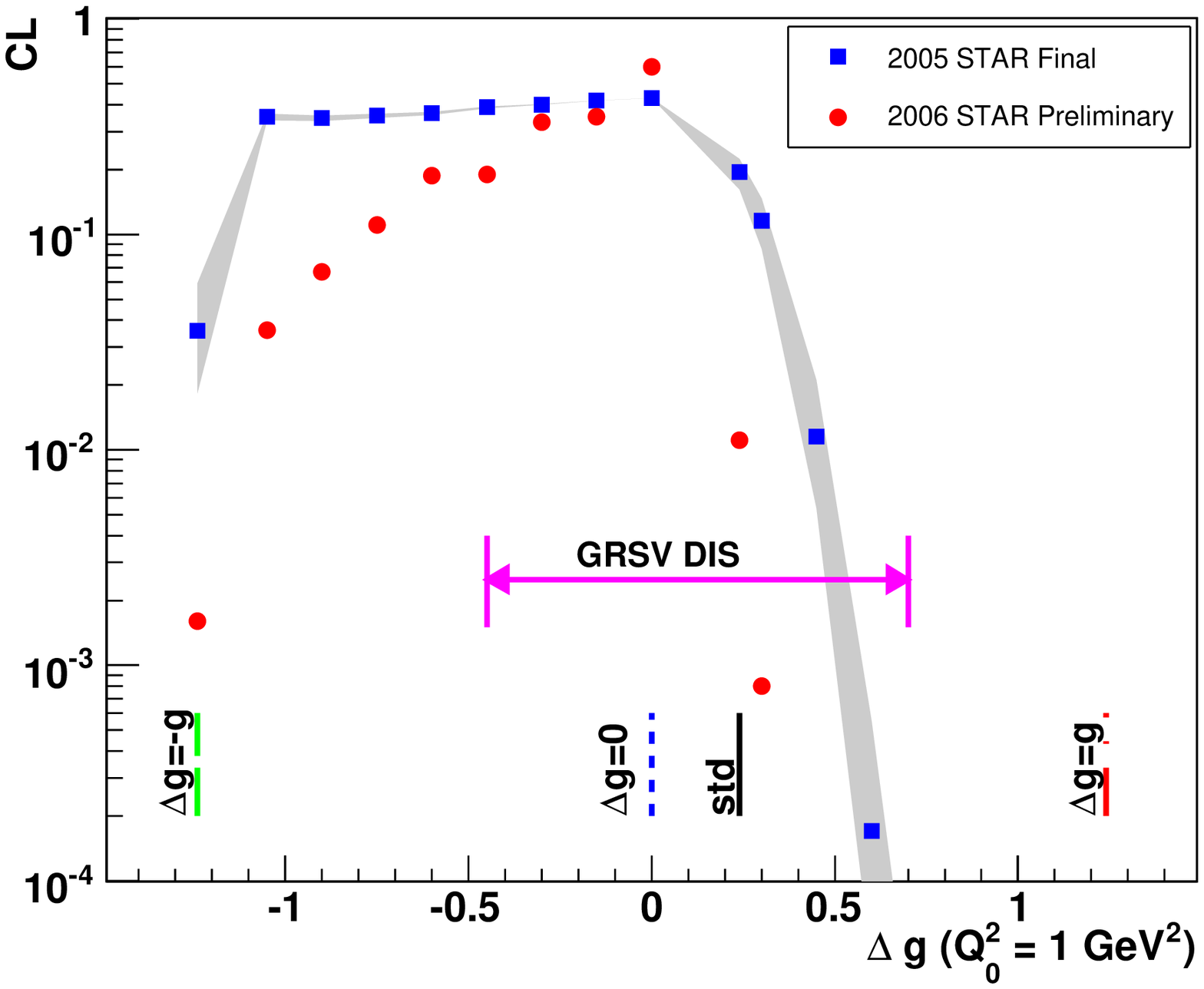}
\caption{\label{fig:aLL06nCL06}\small Left panel: Preliminary 2006 $A_{LL}$
 for inclusive jet production at $\sqrt{s} = 200$ GeV versus jet $p_T$,
 where the error bars are statistical and the gray bands indicate the
 systematic uncertainties. Right panel: Confidence level calculations of
 various predictions in the GRSV framework representing the $A_{LL}$ results
 from 2005 and 2006.}
\end{figure}
  The left panel of Figure~\ref{fig:aLL06nCL06} shows preliminary 2006
 $A_{LL}$ versus jet $p_T$ corrected for detector response. The points with
 error bars are the data with statistical uncertainties. As for the gray
 bands, the height indicates the systematic uncertainties on $A_{LL}$ while
 the width indicates the systematic uncertainties on $p_T$.
 The left panel of Figure~\ref{fig:aLL06nCL06} also shows NLO pQCD calculations
 which incorporate different
 scenarios for $\Delta g(x)$, including the best global fit to the inclusive
 DIS data (std)~\cite{PRD63_094005}.  The other three curves span the full physically allowed
 region from maximally positive gluon polarization ($\Delta g(x) = g(x)$) at the input scale to
 maximally negative gluon polarization ($\Delta g(x) = -g(x)$) and passing
 through zero gluon polarization ($\Delta g(x) = 0$).
 In addition, it includes a prediction derived from GS-C which has a node near $x \sim 0.1$
 as described earlier.
   The statistical uncertainties in the 2006 $A_{LL}$ measurements at high
 $p_T$ are a factor of 3 to 4 smaller than they were in the 2005
 data~\cite{PRL100_232003}, which leads to significantly more stringent
 constraints on gluon polarization models as illustrated in the right panel
 of Figure~\ref{fig:aLL06nCL06}.
 It shows the confidence levels of global analysis predictions within the
 framework of GRSV~\cite{PRD63_094005,MSnWV} representing the $A_{LL}$ results from both runs 2005 and 2006.

 These inclusive jet asymmetries were included in a recent ``global'' NLO
 analysis, DSSV~\cite{PRL101_072001}, which is the first analysis to include
 inclusive DIS, semi-inclusive DIS, and RHIC pp collision data together.
 The result of this analysis showed that $\Delta g(x,Q^2)$ is small
 in the accessible range of momentum fraction, with a possible node in the
 distribution near $x \sim 0.1$, basically evolving away at higher scales
 with the opposite phase from GS-C.
   A study of the DSSV $\chi^2$ profile and the partial contributions
 $\Delta \chi^2$ of the individual data sets shows that these STAR data
 provide the strongest limits on negative gluon polarization over the range
 $0.05 < x < 0.2$ and also contribute significantly to the limits on positive
 gluon polarization over the same range.

  In summary, we reported on $A_{LL}$ measurements from inclusive jet data
 from 2005 and 2006. These results provide significant constraints on the
 gluon spin contribution to the proton spin when compared to NLO pQCD
 calculations. They play a significant role in the first global NLO analysis
 to consider DIS, SIDIS and RHIC data together.
 In addition, we reported a preliminary $A_N$ measurement for inclusive
 jets at mid-rapidity.  It is statistically consistent with zero.




\end{document}